# Design and control of three-dimensional topological magnetic fields using interwoven helical nanostructures

J Fullerton[1*], C Phatak[1,2**]

[1]Materials Science Division, Argonne National Laboratory, Illinois, USA
[2]Department of Materials Science and Engineering, Northwestern University, Evanston, Illinois, USA

Email: *jfullerton@anl.gov; **cd@anl.gov

**Abstract**
Three-dimensional magnetic nanostructures are an emerging platform capable of creating complex topological magnetic fields. The control of localized nanoscale magnetic fields is seen to be of importance for diverse areas from bio-applications such as drug delivery, nanoscale magnetic resonance imaging, as well as condensed matter physics such as particle trapping, and controlling Majorana fermions for quantum computing. Three-dimensional geometric curvature, confinement and proximity can create tailormade spin textures not possible in two-dimensions. The control of magnetization afforded here can allow the formation of unique and reconfigurable stray field patterns and topologies. Here, we report the creation of reconfigurable 3D topological magnetic field textures induced by a single interwoven 3D nanostructure and an applied field protocol. These field textures emerge due to distinct types of domain walls formed in this structure and lead to the creation of an anti-vortex field, a hexapole cusp and a 3D skyrmion field tube of mixed chirality. Our results therefore show a key step towards the design and control of topological magnetic fields on the nanoscale.

## Introduction

Recent advances in nanotechnology increasingly show emphasis shifting away from conventional planar configurations towards more elaborate knotted or three-dimensional (3D) structures [1-3]. In nanomagnetism, this has seen the realization of complex topological spin states such as hopfions, skyrmion tubes and braids, and vortex rings connected by magnetic singularities [3-6]. Similar observations have been made across condensed matter physics, including the formation of optical, plasmonic and ferroelectric skyrmions [7-10]. Correspondingly, there are great potential advantages in understanding the formation of topological electromagnetic structures for applications such as data storage or sensor devices [1, 7]. Included in this is controlling the form of nanoscale magnetic fields in free space [11-12].

The ability to control and manipulate magnetic fields is crucial in many areas of physics and is of common use on the macroscale; from fusion reactors to particle accelerators [13-16]. Here, twisted magnetic fields allow the flow of particles to be directed and controlled, and field cusps allow the confinement of particles or plasma [17-19]. Comparably, the control of localized magnetic fields on the micro/nanoscale has large promise for applications in pinning ultracold atoms, drug delivery, control of nano robots, manipulation of electron and optical beams, and control of Majorana fermions [20-25]. Patterned 2D nanomagnetic structures such as shaped nano islands and artificial spin ice have been observed to create localized nanoscale magnetic stray fields [26-32]. Beyond this, it has recently been shown that 3D magnetic nanostructures offer a promising platform to create even more complex topological stray field textures [11-12, 33-34]. Intertwined double helices can create confined anti-vortex fields [11]. Alternatively, 3D tetrapod and gyroid structures can form chiral and twisted magnetic fields with broken symmetry [12, 34]. Focused electron beam induced deposition (FEBID) enables us a route towards the design and realization of 3D magnetic nanostructures through utilizing software like f3ast [11, 35-37]. These nanostructures can possess intricate spin textures that can be controlled by geometric patterning and reconfigured through applied magnetic field protocols [11, 36-37].

In this work, we will discuss the controlled formation of reconfigurable topological stray field textures through altering the magnetization in a 3D nanostructure. We use FEBID to fabricate nanostructures formed by several interwoven helical nanowires. By connecting multiple nanowires together, we can create a series of wire junctions reminiscent of tetrapod and 3D artificial spin ice structures. However, the helical nature of the interwoven nanowires introduces an intrinsic geometric curvature and chirality. We observe that the introduction of 3D curvature here is vital for the formation of unique and complex stray fields. We subsequently characterize this emergent stray field with off-axis electron holography. Micromagnetic simulations allow us to finely analyze and understand the form of the stray field and correlate them with the local spin configurations that induce them. The application of a global magnetic field allows us the ability to reconfigure the magnetization of the nanostructure to form multi-domain or single domain states. Depending on the magnetization configuration, corresponding distributions of magnetic charges will arise throughout the nanostructure, leading to unique forms of the emanating stray field. These results show how 3D nanostructures could be utilized in the creation of topological magnetic field textures and provide insight into how they can be controllably reconfigured and designed.

Results

Through focused electron beam induced deposition (FEBID), we can design and fabricate functional 3D magnetic nanostructures [11, 35-37]. This work concentrates on a nanostructure formed of four helical nanowires (with matching chirality) interwoven like a chain link fence; a schematic is shown in Fig. 1a and an experimentally realized structure is shown in Fig. 1b. The interwoven nanowires form a series of wire junctions (excerpt shown in Fig. 1c) which act as nucleation sites for magnetic domain walls following the application of a saturating magnetic field. Figures 1d-f show micromagnetic simulations of three possible magnetic configurations in a wire junction after the application and removal of a global magnetic field along the x, y and z-axes, respectively. In each case, applying a field in this manner results in a "2-in/2-out" spin configuration with the magnetization in the wire junction roughly aligned with the direction of the applied field. The application of a field along the x-axis creates a transverse-like domain wall, Fig 1d, a field along the y-axis creates a helical domain wall with anti-vortex surface textures [12], Fig 1e, and finally, a field along the z-axis creates a single domain axial state, Fig. 1f. Experimental magnetic induction images reconstructed using the transport of intensity method of the three magnetic states described above are contained in supplementary information section 1. It is expected that other, metastable junction configurations are possible, for example, a "3-in/1-out" spin arrangement. However, for the purposes of this work, we will only consider the three "2-in/2-out" magnetic configurations as they can be universally set throughout the entire interwoven nanostructure (Figs. 1a and b) with a global magnetic field along a chosen axis.

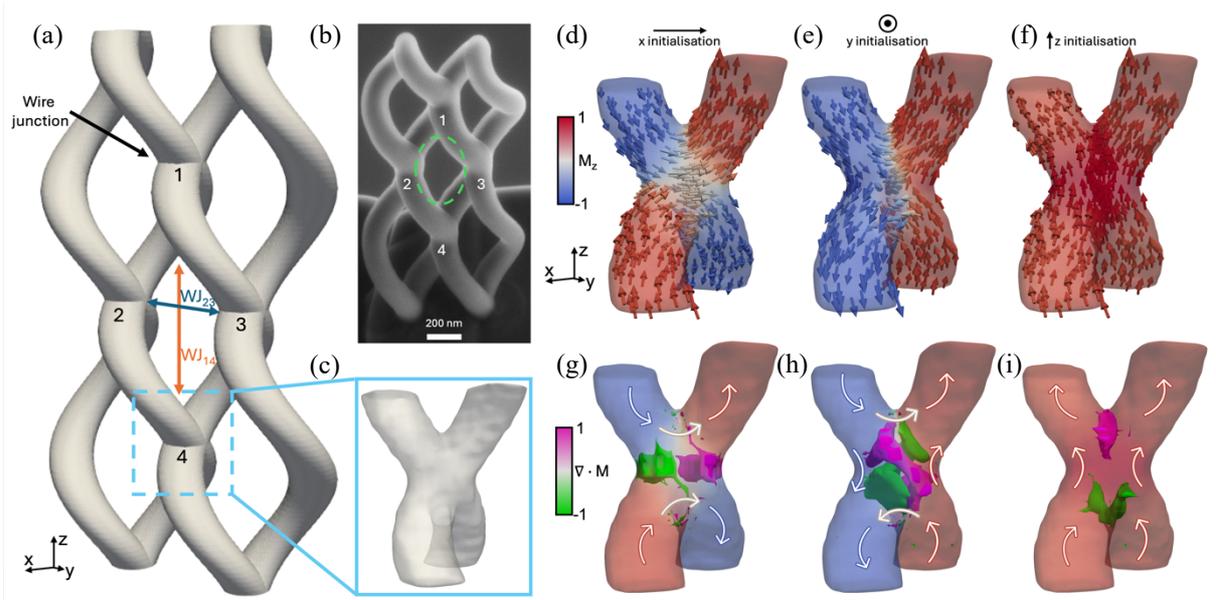

Fig 1: Interwoven nanowire fabrication and tetrapod states. (a) 3D CAD model of an interwoven nanostructure formed of four helical nanowires. The junctions between the wires form multiple tetrapod junctions throughout the structure. (b) An example interwoven cobalt nanostructure fabricated by FEBID. (c) Depiction of the geometry of a single tetrapod junction. (d-f) Micromagnetic simulations colored by the $M_z$ component of three possible "2-in/2-out" magnetic states in the tetrapods formed by applying a saturating magnetic field along the x-axis (d), y-axis (e) and z-axis (f). (g-i) Micromagnetic simulations showing the distribution of magnetic volume charges of the three possible "2-in/2-out" magnetic states in the tetrapods formed by applying a saturating magnetic field along the x-axis (g), y-axis (h) and z-axis (i). Pink and green isosurfaces correspond to areas of maximum and minimum divergence of the magnetization to visualize the distribution of magnetic charge formed by each state.

The application of a magnetic field not only rearranges the spin configuration at each wire junction, but it also rearranges the distribution of magnetic charge throughout the system [38-39]. By plotting isosurfaces of the maximum and minimum divergence of the magnetization in simulations, we can visualize the regions of positive (pink) and negative (green) magnetic volume charge, as shown for each case in Fig. 1g-i. After applying a field along the x- or z-axis, the magnetization in the wire junction lies along the axis that the field was applied. Therefore, we see C-shaped accumulation of charges, curving around the edges of the junction along the direction of the field that was applied (*i.e.*, left-to-right in Fig. 1g, and bottom-to-top in Fig. 1i). However, in the case of initializing the structure with a field along the y-axis, we see a more complex arrangement of charge. This complexity is caused by the twisting of the helical domain wall due to the 3D curvature of the nanowires. Consequently, the isosurfaces of opposing magnetic charges form interlocking U-shapes orthogonal to each other. We show here that we can control and reconfigure both the magnetization and the distribution of magnetic charge in a wire junction formed by our interwoven nanostructure via a simple external field sequence.

In our full interwoven nanostructure (Figs. 1a/b), we form multiple wire junctions separated by gaps of the order of ~300 nm throughout the structure. As we can controllably initialize the magnetic state of each junction, we can subsequently control the distribution of magnetic charges around each gap. Consequently, the magnetic charges act like magnetic poles and induce complex, topological stray field textures in between them. For the rest of the main text, we focus on magnetic fields created in the central gap of the structure (shown by the green oval in fig. 1b). The central gap is surrounded by four wire junctions labelled as 1, 2, 3 and 4 in fig. 1a and b. However, this isn't the only location in the structure capable of supporting complex stray field patterns.

Hence, in supporting information section 2 we also highlight specific cases in other gaps formed by the nanostructure. Below, we discuss the form of the stray field after a saturating global field has been applied and removed along each axis in turn. To verify the presence of these topological field patterns, we use electron holography to map the stray field of our experimentally patterned structure and compare to simulated results [40-43]. The electron holography images shown throughout this work are 2D projections along the electron beam-direction (*i.e.*, the x-axis) of the 3D magnetic induction created by the interwoven nanostructure. As such, there may be slight differences in appearance of the experimental and simulated holography images due to sample tilt/changes in domain wall location, etc. Additionally, it should be noted that due to carbon accumulation during holography measurements, multiple experimental structures were imaged and are presented throughout this paper (example structures are shown in supporting information section 1). However, each structure used here was fabricated in the same manner and only slight variations in aspect ratio are present. Specifically, there is a variation of the ratio of the distance between wire junctions 1 and 4 ($WJ_{14}$) and the distance between wire junctions 2 and 3 ($WJ_{23}$), as defined in Fig. 1a, of between 1:2 to 1:3. Simulations confirm that this variation does not impact the form of the stray field, just slightly the appearance of the electron holography image. Therefore, due to the above factors, simulated holography images will vary in aspect ratio, etc. to match the experimental images the closest. In supporting information section 3, we discuss the impact on the stray field of increasing the $\frac{WJ_{14}}{WJ_{23}}$ ratio to 1:4 in the case of an axial saturating field.

First, we consider the form of the stray field emerging from the interwoven nanostructure after applying a global field along the x-axis (fig. 2a). Figure 2b shows an experimental stray field map in the central nanostructure gap, obtained through electron holography after applying a large field into the plane of the interwoven nanostructure. Correspondingly, fig. 2c shows the simulated electron holography image from the magnetic state shown in fig. 2d. In both, we observe a similar stray field map featuring field lines similar to an anti-vortex-like structure towards the bottom of the gap. In fig. 2d, we show a micromagnetic simulation of the nanostructure where each wire junction has formed transverse domain walls along the x-axis. This creates accumulations of positive and negative magnetic charges at the front and the back of the structure (denoted by '+' and '−' signs in Fig. 2d). The resulting stray field from this distribution of charges in our central gap appears to take the form like that of an anti-vortex (Fig. 2d). However, there is a more 3D structure to the field. Figures 2e-g show yz slices of the stray field at three points along the x-axis. We see that in front of and behind the nanostructure (Figs. 2e and g, respectively), the field forms anti-vortices of opposite chirality (both with polarity along the x-axis). In the center of the gap (Fig. 2f), the stray field is more uniform and points along the x-axis. It should be noted that the field here points against the direction of the initial global field.

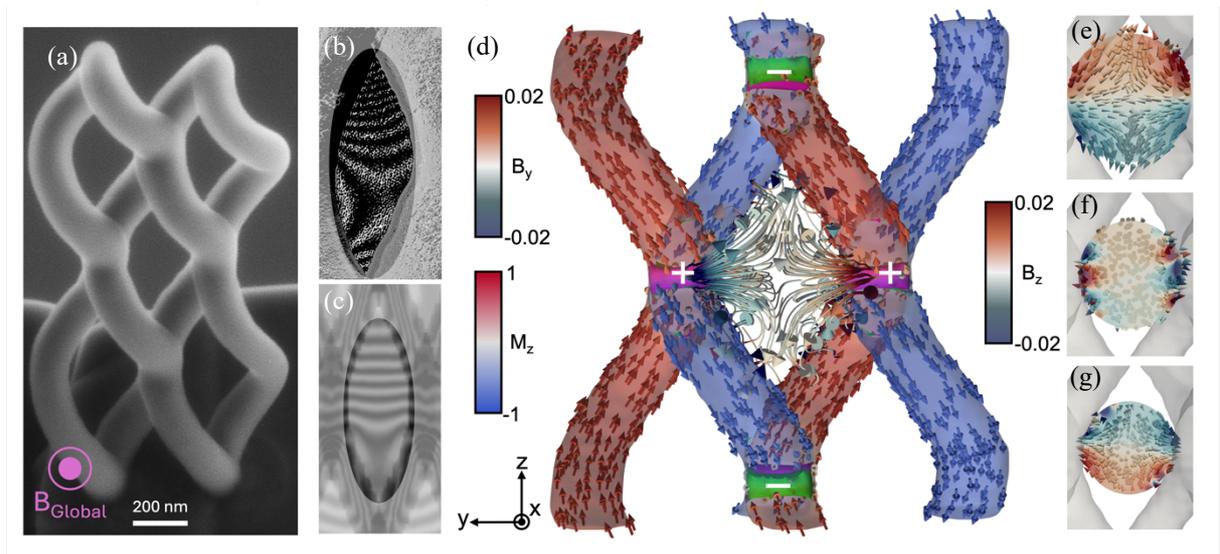

Fig 2: Simulated magnetic stray field and holography imaging of interwoven nanostructure after x-axis field initialization. (a) SEM image of interwoven nanostructure showing direction of applied global magnetic field. (b) Experimental electron holography image in the center gap after applying a strong magnetic field along the x-axis of the nanostructure. (c) Simulated electron holography image in the center gap of the nanostructure with magnetization as shown in part (d). (d) Micromagnetic simulation of magnetization in an interwoven nanostructure after x-axis initialization and resulting stray field in center gap. Magnetization is colored by the $M_z$ component while the stray field is colored by the $B_y$ component and pink and green isosurfaces represent areas of maximum and minimum divergence created by the domain walls. (e-g) Slices in the yz plane showing the stray field configuration colored by the $B_z$ component in front of (e), in-between (f), and behind the center gap (g).

In Fig. 3 we describe the form of the stray field after applying a saturating field along the z-axis (fig. 3a) and creating a single domain state. Figures 3b and c show simulated and experimental electron holography images of the stray field after a global z-axis field, respectively. Here, both display similar patterns to a multipole field cusp. In fig. 3d we show a micromagnetic simulation of the interwoven nanostructure after a saturating axial field. Here, the converging and diverging geometry of our nanostructure creates six regions of alternating positive and negative charges around the central gap of the nanostructure. These six charges create a stray field reminiscent of an anti-vortex shape but with some differences. Here, the six charges lead to the field to have six lobes (as opposed to four in a standard anti-vortex). Figure 3e shows a plot of the field in the center of the gap to highlight the six-fold structure. Furthermore, due to this arrangement, the magnitude of the field directly in the center goes towards zero. Hence, we create a field texture comparable to a multipole toroidal cusp utilized in plasma physics, but now on the nanoscale [17]. As the field texture in this case is dependent on magnetic charges from all four wire junctions around the central gap, its specific structure is dependent on the relative distance between all charges. As such, in supporting information section 3 we detail how increasing the $\frac{WJ_{14}}{WJ_{23}}$ ratio to 1:4 can alter the field creating linked four-fold multipole cusps.

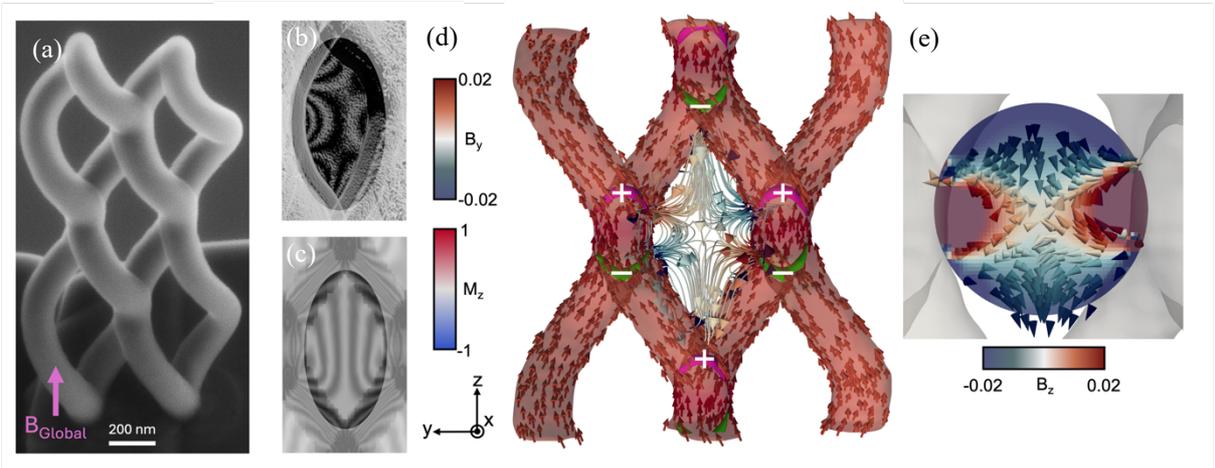

Fig 3: Simulated magnetic stray field and holography imaging of interwoven nanostructure after z-axis field initialization. (a) SEM image of interwoven nanostructure showing direction of applied global magnetic field. (b) Experimental electron holography image in the center gap after applying a strong magnetic field along the z-axis of the nanostructure. (c) Simulated electron holography image in the center gap of the nanostructure with magnetization as shown in part (d). (d) Micromagnetic simulation of magnetization in an interwoven nanostructure after z-axis initialization and resulting stray field in center gap. Magnetization is colored by the $M_z$ component while the stray field is colored by the $B_y$ component and pink and green isosurfaces represent maximum and minimum divergence created by the domain walls. (e) A slice in the yz plane showing the stray field configuration colored by the $B_z$ component in the center gap.

The two stray field configurations described above emerge from similar C-shaped arrangements of magnetic charges either formed along the x- or z-axis. However, as described in Fig. 1e and h, the application of a global field along the y-axis (fig. 4a), combined with the 3D curvature of the nanostructure, leads to a much more intricate arrangement of magnetic charge. Correspondingly, this creates a more complex stray field texture. In Fig. 4 we discuss experimental observation and simulations of this topological 3D stray field texture, which takes the form reminiscent of a skyrmion tube with mixed chirality. We see a good agreement between the experimental and simulated holography images in Fig. 4b and c, respectively, as both show a similar double anti-vortex-like structure appearing in the central gap. In Fig. 4d, we show a micromagnetic simulation of the stray field emerging from two helical domain walls on the left and right of our interwoven nanostructure. The complex arrangement of magnetic charge creates a central core of field lines (red/orange) emitting from the positive (pink) charges on the right-hand side, towards the negative (green) charge on the left. This core is surrounded by an outer shell of field lines (blue) in the opposite direction emitting from the opposite charges in each wire junction. Therefore, giving the emergent field the appearance of a confined tube. In supplementary information section 4 we include a schematic description of the formation of this stray field texture to aid understanding.

To further understand the 3D form of this field configuration, in Fig. 4e we plot xz cross-sections of the stray field in the central gap at three points along the y-axis. Complementary plots of the magnetic field components and magnitude along the diameter of each cross-section are shown in Fig. 4f. In the left- and right-hand side cross-sections, we observe that the stray field components take the form resembling that of a Néel skyrmion but with opposite chiralities (*i.e.*, an anticlockwise rotation of components from left-to-right in the left-hand cross-section, and a clockwise rotation in the right-hand cross-section). This is also highlighted by the reversed profile of the $B_x$ component between the left- and right-hand plots in Fig.4 f, which shows the rotation of the "domain wall". In the central cross-section, we observe the same inner core and outer shell field directions. However, in the "domain wall" between the core and shell, all components of the field, and hence the magnitude, go towards zero. Thus, creating a ring cusp mediating the apparent opposite chiralities of magnetic field. Figures 4g and h show contour plots of the magnetic field skyrmion to understand how the ring cusp impacts its 3D structure. In Fig. 4g we plot contours of the $B_y$ component and the overall field magnitude sliced in half for visualization. As before, we observe the inner $+B_y$ core (red/orange) in the middle and the outer $-B_y$ shell (blue) on the top and bottom. There is clear constriction of the $+B_y$ core towards the center of the plot which coincides with a reduction in the magnitude of the overall field where the ring cusp is located (see white-green-black colors). Fig. 4h shows the isosurface where $B_y = 0$, colored by the azimuthal component (i.e., $B_x$ and $B_z$). Here, we further emphasize the resemblance to a Néel skyrmion tube with opposite chiralities on the left- and right-hand sides. Here, we also plot a black isosurface where the magnitude of the stray field goes to 0. This zero field isosurface appears as ring around the center and mediates this apparent reversal of chiralities.

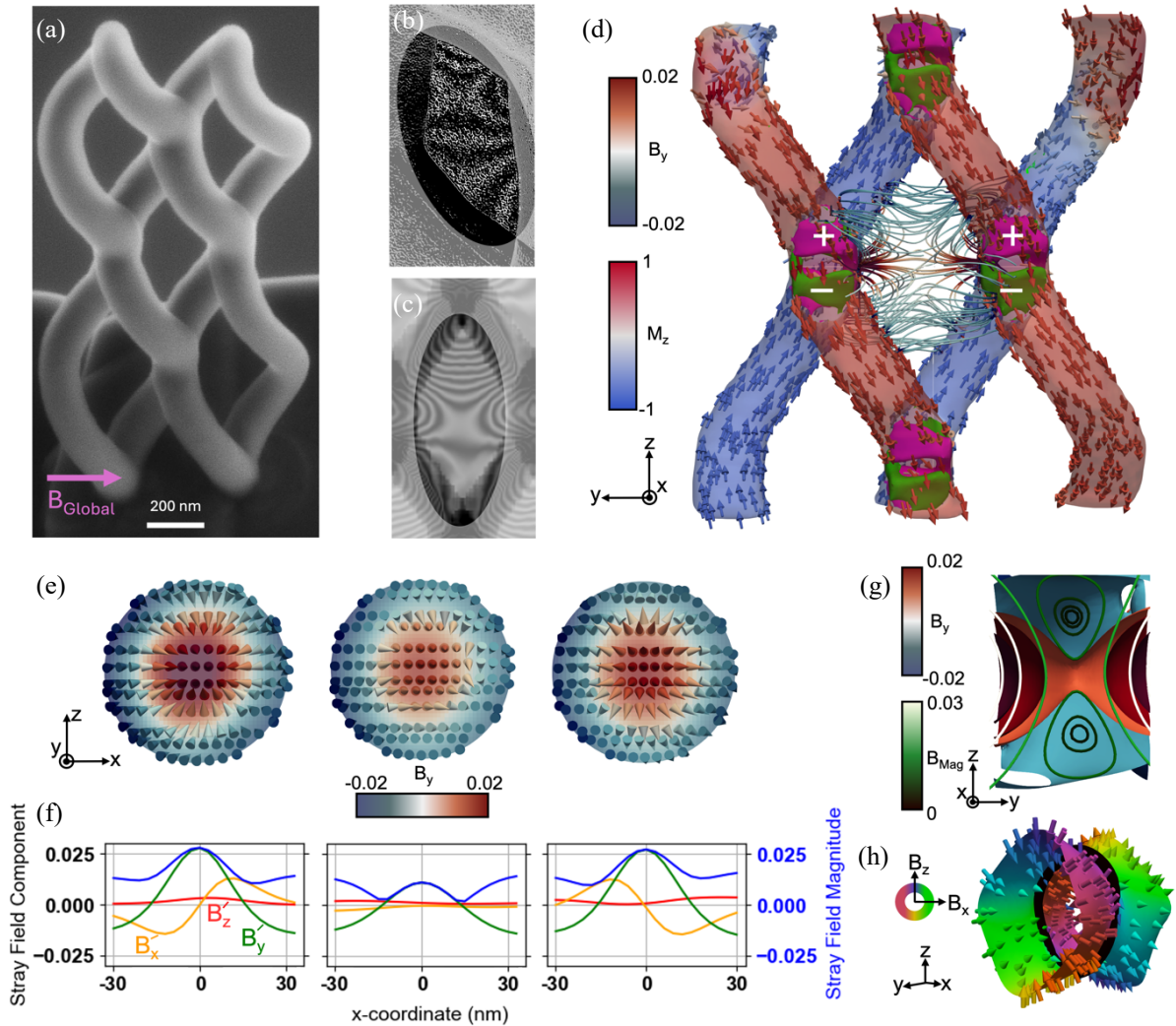

Fig 4: Simulated magnetic stray field and holography imaging of interwoven nanostructure after y-axis field initialization. (a) SEM image of interwoven nanostructure showing direction of applied global magnetic field. (b) Experimental electron holography image in the center gap after applying a strong magnetic field along the y-axis of the nanostructure. (c) Simulated electron holography image in the center gap of the nanostructure with magnetization as shown in part (d). (d) Micromagnetic simulation of magnetization in an interwoven nanostructure after y-axis initialization and resulting stray field in center gap. Magnetization is colored by the $M_z$ component while the stray field is colored by the $B_y$ component and pink and green isosurfaces represent maximum and minimum divergence created by the domain walls. (e) Slices in the xz plane at three-points along the y-axis showing the stray field configuration colored by the $B_y$ component on the left, middle and right hand side of the center gap. (f) Profiles of the stray field components and magnitude along the diameter of the cross-sections shown in part (e). (g) Contour plots of the $B_y$ stray field component (blue to red) and field magnitude (black to green to white). (h) Profile of the isosurface where $B_y = 0$, colored by the $B_x$ and $B_z$ components showing opposite chirality on each side with a black ring representing the isosurface where $B_{Mag} \to 0$.

## Discussion

By utilizing controlled 3D-nanoprinting and a simple magnetic field protocol we have demonstrated the ability to create and reconfigure localized topological 3D magnetic fields on the nanoscale. By designing wire junctions in interwoven curved nanostructure, we can initialize multiple, distinct domain walls with an applied field. We therefore induce arrangements of magnetic charge that lead to the emergence of topological stray fields due to the designed geometric curvature and proximity. We show the formation of an anti-vortex field, a hexapole field cusp and a complex 3D skyrmion tube like structure featuring mixed chiralities and a ring field cusp. The skyrmion tube-like stray field configuration described above exhibits a complex structure in 3D, which arises directly from the 3D curvature of our interwoven nanostructure. In supporting information section 5, we consider the emergent stray field from an analogous 2D, planar magnetic lattice. In the 2D case, the full 3D "skyrmion tube" structure is lost, and instead a planar double anti-vortex field texture is formed. Therefore, the 3D geometry and curvature shown in our interwoven nanostructure is vital to creating the complex topology we discuss. Our results further emphasize that 3D nanopatterning offers a unique platform for the design and control of magnetization and resulting stray fields. This work demonstrates potential routes for creating bespoke magnetic fields, potentially useful for a variety of nanotechnology applications. Specifically, the field cusps and null points could allow the trapping or guiding of particles and atoms. Alternatively, more uniform fields could direct nanorobots. Moreover, as greater research emphasis is placed on fully 3D topological structures across physics, it follows that we may require 3D nanostructures to induce confined rings and hopfion-like magnetic field textures.

## Methods

*Nanofabrication and simulation*

The interwoven cobalt nanostructures were fabricated using focused electron beam-induced deposition in an FEI Nova dual beam FIB/SEM. The precursor gas used was $Co_2(CO)_8$ with a working pressure of $3\times10^{-6}$ Torr, accelerating voltage of 5 kV and a beam current of 96 pA. For the purposes of this work, we fabricate FEBID nanowires of diameter up to ~100-140 nm. While this makes imaging of the magnetic textures more challenging, it allows a strong magnetic moment and clear detection of the stray field patterns. The nanohelices were grown directly onto copper lift-out TEM grids before being transferred to the TEM.

The nanostructures were designed using the 3D CAD software OpenSCAD. From this stl files were exported and used for fabrication and simulations. To fabricate the samples, the software f3ast was used to create stream files that contained a series of x and y pixel coordinates and dwell times for each point [35]. For simulation, the computer models were discretized using the Voxelize function in PyVista. A model with voxels of size 3x3x3 nm was created and inputted into MuMax3 to simulate the magnetization of the nanostructure [44]. A saturation magnetization of 900 kAm$^{-1}$ and an exchange stiffness of 10$^{-11}$ Jm$^{-1}$ was used to be representative of FEBID cobalt.

*Transmission electron microscopy*

The off-axis electron holography (and magnetic phase retrieval) was performed in a JEOL 2100F LTEM instrument operating at 200 kV with an imaging Cs corrector. The samples sit in a low field environment to avoid affecting the initialized magnetic state. We used the open source PyLorentz code developed in our group to simulate the electron holography images from the output of the MuMax3 simulations [45]. The experimental retrieval and simulation of the magnetic phase was performed using the transport of intensity equation method also using PyLorentz. Both the simulated and experimental holography field maps are generated by multiplying the cosine of the calculated magnetic phase by a factor of five. The resulting black and white field lines are representative of the projected magnetic induction.

## Acknowledgements

This work was supported by the U.S. Department of Energy, Office of Science, Office of Basic Energy Sciences, Materials Sciences and Engineering Division. Use of the Center for Nanoscale Materials, an Office of Science user facility, was supported by the U.S. Department of Energy, Office of Science, Office of Basic Energy Sciences, under Contract No. DE-AC02-06CH11357. The authors would like to acknowledge Dave Czaplewski for technical support at the Center for Nanoscale Materials. We would also like to thank Trevor Almeida, Kayla Fallon and William Smith at the Kelvin Nanocharacterisation center at the University of Glasgow for their support as preliminary structures were fabricated there.

## References

[1] A. Fernández-Pacheco, et. al., Three-dimensional nanomagnetism, *Nat. Commun.*, **8**, 15756 (2017).
[2] Y. Wang, et. al., Free-space direct nanoscale 3D printing of metals and alloys enabled by two-photon decomposition and ultrafast optical trapping, *Nat. Mater.* (2024).
[3] F. Zheng, et. al., Hopfion rings in a cubic chiral magnet, *Nature*, **623**, 718 (2023).
[4] D. Wolf., et. al., Unveiling the three-dimensional magnetic texture of skyrmion tubes, *Nat. Nanotechnol.*, **17**, 250 (2022).
[5] F. Zheng, et. al., Magnetic skyrmion braids, *Nat. Commun.*, **12**, 5316 (2021).
[6] C. Donnelly, et. al., Experimental observation of vortex rings in a bulk magnet, *Nat. Phys.*, **17**, 316 (2021).
[7] Y. Shen, et. al., Optical skyrmions and other topological quasiparticles of light, *Nat. Photon.*, **18**, 15 (2024).
[8] Y. Shen., et. al., Generation of optical skyrmions with tunable topological textures, *ACS Photonics*, **9**, 296 (2022).
[9] ZL. Deng, et. al., Observation of localized magnetic plasmon skyrmions, *Nat. Commun.*, **13**, 8 (2022).
[10] S. Das, et. al., Observation of room-temperature polar skyrmions, *Nature*, **568**, 368 (2019).

[11] C. Donnelly, et. al., Complex free-space magnetic field textures induced by three-dimensional magnetic nanostructures, *Nat. Nanotechnol.*, **17**, 136 (2022).
[12] O. M. Volkov, et. al., Three-dimensional magnetic nanotextures with high-order vorticity in soft magnetic wireframes, *Nat. Commun.*, **15**, 2193 (2024).
[13] J. Ongena, et. al., Magnetic-confinement fusion, *Nat. Phys.*, **12**, 398 (2016).
[14] M. Johansson, et. al., Magnet design for a low-emittance storage ring, *J. Synchrotron Radiat.*, **21**, 884 (2014).
[15] N. Bort-Soldevila, et. al., Complete and robust magnetic field confinement by superconductors in fusion magnets, *Sci. Rep.*, **14**, 3653 (2024).
[16] T. Sunn Pedersen, et. al., Confirmation of the topology of the Wendelstein 7-X magnetic field to better than 1:100,000, *Nat. Commun.*, **7**, 13493 (2016).
[17] M. Landreman, et. al., Magnetic fields with precise quasisymmetry for plasma confinement, *Phys. Rev. Lett.*, **128**, 035001 (2022).
[18] I. J. D. Craig, et. al., Current singularities in line-tied three-dimensional magnetic fields, *APJ.*, **788** (2014)
[19] P. Wyper, et. al., Torsional magnetic reconnection at three dimensional null points: A phenomenological study, *Phys. Plasmas*, **17**, 092902 (2010).
[20] A. D. West, et al., Realization of the manipulation of ultracold atoms with a reconfigurable nanomagnetic system of domain walls, *Nano Lett.*, **12**, 4065 (2012).
[21] B. Wang, et. al., Trends in micro-/nanorobotics: Materials development, actuation, localization, and system integration for biomedical applications, *Adv. Mater.*, **33**, 2002047 (2021).
[22] R. Huber, et. al., Tailoring electron beams with high-frequency self-assembled magnetic charged particle micro optics, *Nat. Commun.*, **13**, 3220 (2022).
[23] H. Wang, et. al., Three-degrees-of-freedom orientation manipulation of small untethered robots with a single anisotropic soft magnet, *Nat. Commun.*, **14**, 7491 (2023).
[24] K. Qu, et. al., Plasma q-plate for generation and manipulation of intense optical vortices, *Phys. Rev. E*, **96**, 053207 (2017).
[25] M. Kjaergaard, et. al., Majorana fermions in superconducting nanowires without spin-orbit coupling, *Phys. Rev. B*, **85**, 020503 (2012).
[26] S. Gliga, et. al., Emergent dynamic chirality in a thermally driven artificial spin ratchet, *Nat. Mater.*, **16**, 1106 (2017).
[27] S. Gliga, et. al., Switching a magnetic antivortex core with ultrashort field pulses, *J. Appl. Phys.*, **103**, 07B115 (2008).
[28] D. Kumar, et al., Magnetic vortex based transistor operations, *Sci. Rep.*, **4**, 4108 (2014).
[29] L. Rondin, et. al., Stray-field imaging of magnetic vortices with a single diamond spin, *Nat. Commun.*, **4**, 2279 (2013).
[30] S. Jenkins, et. al., Magnetic stray fields in nanoscale magnetic tunnel junctions, *J. Phys. D: Appl. Phys.*, **53**, 044001 (2020).
[31] M. A. Tschudin, et. al., Imaging nanomagnetism and magnetic phase transitions in atomically thin CrSBr, *Nat. Commun.*, **15**, 6005 (2024).
[32] S. Li, et. al., Observation of stacking engineered magnetic phase transitions within moiré supercells of twisted van der Waals magnets, *Nat. Commun.*, **15**, 5712 (2024).
[33] D. Vasyukov, et. al., Imaging stray magnetic field of individual ferromagnetic nanotubes, *Nano Lett.*, **18**, 964 (2018).
[34] L. V. Lich, et. al., Formation and switching of chiral magnetic field textures in three-dimensional gyroid nanostructures, *Acta Mater.*, **249**, 118802 (2023).

[35] L. Skoric, et. al., Layer-by-layer growth of complex-shaped three-dimensional nanostructures with focused electron beams, *Nano Lett*., **20**, 184 (2020).

[36] J. Fullerton, et. al., Understanding the effect of curvature on the magnetization reversal of three-dimensional nanohelices, *Nano. Lett.*, **24**, 2481 (2024).

[37] D. Sanz-Hernández, et. al., Artificial double-helix for geometrical control of magnetic chirality, *ASC Nano.*, **14**, 8084 (2020).

[38] A. S. Arrott, et. al., Using magnetic charge to understand soft-magnetic materials, *AIP Adv.*, **8**, 047301 (2017).

[39] J. Fullerton, et. al., Controlled evolution of three-dimensional magnetic states in strongly coupled cylindrical nanowire pairs, *Nanotechnol.*, **34**, 125301 (2023).

[40] T. Tanigaki, et. al., Magnetic field observations in CoFeB/Ta layers with 0.67-nm resolution be electron holography, *Sci. Rep.*, **7**, 16598 (2017).

[41] J. Llandro, et. al., Visualizing magnetic structure in 3D nanoscale Ni-Fe gyroid networks, *Nano Lett.*, **20**, 3642 (2020).

[42] T. P. Almeida, et. al., Visualized effect of oxidation on magnetic recording fidelity in pseudo-single-domain magnetite particles, *Nat. Commun.*, **5**, 5154 (2014).

[43] V. Boureau, et. al., An electron holography study of perpendicular magnetic tunnel junctions nanostructured by deposition of pre-patterned conducting pillars, *Nanoscale*, **12**, 17312 (2020).

[44] Vansteenkiste A, et. al., The design and verification of MuMax3, *AIP Adv.*, **4,** 107133 (2014).

[45] McCray A R C, et. al., Understanding complex magnetic spin textures with simulation-assisted Lorentz transmission electron microscopy *Phys. Rev. Appl.*, **15,** 044025 (2021).